\documentclass[11pt]{article}

\usepackage[final]{acl}

\usepackage{times}
\usepackage{latexsym}
\usepackage{float}
\usepackage[T1]{fontenc}

\usepackage[utf8]{inputenc}

\usepackage{microtype}
\usepackage{amssymb}
\usepackage{pifont}
\newcommand{\xmark}{\ding{55}}

 \usepackage{microtype}
 \usepackage{url}
\usepackage{graphicx}
\usepackage{amsmath}
\usepackage{amssymb}
\usepackage{booktabs}
\usepackage{multirow}
\usepackage{array}
\usepackage{listings}
\usepackage{xcolor}
\usepackage{url}
\usepackage[breakable]{tcolorbox} 
\usepackage{enumitem} 
\usepackage{graphicx}
\usepackage{makecell}
\title{Evaluating the Expressive Appropriateness of Speech in Rich Contexts}

\author{
 \textbf{Tianrui Wang\textsuperscript{1,2}},
 \textbf{Ziyang Ma\textsuperscript{2,3}},
 \textbf{Yizhou Peng\textsuperscript{2}},
 \textbf{Haoyu Wang\textsuperscript{1}},
 \textbf{Zhikang Niu\textsuperscript{3}},
  \\
 \textbf{Zikang Huang\textsuperscript{1}},
 \textbf{Yihao Wu\textsuperscript{2}},
 \textbf{Yi-Wen Chao\textsuperscript{2}},
 \textbf{Yu Jiang\textsuperscript{1}},
 \textbf{Yuheng Lu\textsuperscript{1}},
 \\
 \textbf{Guanrou Yang\textsuperscript{3}},
 \textbf{Xuanchen Li\textsuperscript{1}},
 \textbf{Hexin Liu\textsuperscript{2}},
 \textbf{Chunyu Qiang\textsuperscript{1,4}},
 \textbf{Cheng Gong\textsuperscript{5}},
 \\
 \textbf{Yifan Yang\textsuperscript{3}},
 \textbf{Tianchi Liu\textsuperscript{6}},
 \textbf{Junyu Wang\textsuperscript{1}},
 \textbf{Nana Hou\textsuperscript{2}},
 \textbf{Meng Ge\textsuperscript{1}},
 \\
 \textbf{Fuming You\textsuperscript{7}},
 \textbf{Wei Yang\textsuperscript{7}},
 \textbf{Zhongqian Sun\textsuperscript{7}},
 \textbf{Haifeng Hu\textsuperscript{7}},
 \textbf{Xiaobao Wang\textsuperscript{1}\thanks{Longbiao Wang is the primary corresponding author and Xiaobao Wang is the co-corresponding author.}},
  \\
 \textbf{Eng Siong Chng\textsuperscript{2}},
 \textbf{Xie Chen\textsuperscript{3}},
 \textbf{Longbiao Wang\textsuperscript{1}\footnotemark[1]},
 \textbf{Jianwu Dang\textsuperscript{1}}
\\
 \textsuperscript{1}Tianjin Key Laboratory of Cognitive Computing and Application, School of Artificial\\Intelligence, Tianjin University,
 \textsuperscript{2}Nanyang Technological University,\\
 \textsuperscript{3}Shanghai Jiao Tong University,
 \textsuperscript{4}Kuaishou Technology,
 \textsuperscript{5}TeleAI, China Telecom,\\
 \textsuperscript{6}National University of Singapore,
 \textsuperscript{7}Tencent
\\
 \small{
   \textbf{Correspondence:} \href{mailto:longbiao_wang@tju.edu.cn}{\{longbiao\_wang, wangxiaobao\}@tju.edu.cn}
 }
}

\begin{document}
\maketitle
\begin{abstract}
Evaluating expressive speech remains challenging, as existing methods mainly assess emotional intensity and overlook whether a speech sample is expressively appropriate for its contextual setting. This limitation hinders reliable evaluation of speech systems used in narrative-driven and interactive applications, such as audiobooks and conversational agents. 
We introduce \textbf{CEAEval}, a \textbf{C}ontext-rich framework for \textbf{Eval}uating \textbf{E}xpressive \textbf{A}ppropriateness in speech, which assesses whether a speech sample expressively aligns with the underlying communicative intent implied by its discourse-level narrative context.
To support this task, we construct CEAEval-D, the first context-rich speech dataset with real human performances in Mandarin conversational speech, providing narrative descriptions together with fifteen dimensions of human annotations covering expressive attributes and expressive appropriateness.
We further develop CEAEval-M, a model that integrates knowledge distillation, planner-based multi-model collaboration, adaptive audio attention bias, and reinforcement learning to perform context-rich expressive appropriateness evaluation. Experiments on a human-annotated test set demonstrate that CEAEval-M substantially outperforms existing speech evaluation and analysis systems. 
\end{abstract}

\section{Introduction}
\vspace{-0.1cm}
Automatic speech evaluation has long supported tasks such as data filtering and model optimization \cite{crowdmos}. 
With the rapid deployment of speech dialogue systems \cite{wavchat} and audiobook-generation models \cite{audiobook}, the expressive quality has become a critical factor shaping user experience \cite{cong2021controllable}. 

\begin{figure}[t]
  \centering
  \includegraphics[width=0.49\textwidth]{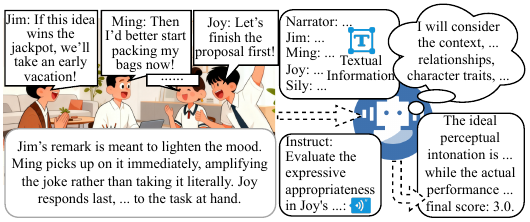}
  \vspace{-0.6cm}
  \caption{Overview of the proposed context-rich expressive appropriateness evaluation task. The dialogue example is shown in English for illustrative purposes.}
\vspace{-0.2cm}
  \label{fig:context-example}
\end{figure}

\begin{table*}[t]
\centering
\small
\setlength{\tabcolsep}{2pt}
\begin{tabular}{lccccccc}
\toprule
Benchmark / Work
& \makecell{Real\\Speech}
& \makecell{Real\\Context}
& \makecell{Long-range\\Context}
& \makecell{Multiple\\Turn}
& \makecell{CoT-based\\Reasoning}
& \makecell{Number of\\Annotation\\Dimensions}
& \makecell{Task Focus}
\\
\midrule

WavReward \cite{ji2025wavreward}
& \xmark & \xmark & \xmark & \checkmark  & \checkmark & 1 & Spoken Dialogue Quality\\

SpeechJudge \cite{zhang2025speechjudge}
& \xmark & \xmark & \xmark & \xmark  & \checkmark & 2 & Speech Naturalness  \\

Speech-DRAME \cite{speechdrame}
&\checkmark & \xmark & \xmark & \checkmark  & \xmark & 13 & Role-play Interaction \\

SpeechRole \cite{jiang2025speechrole}
& \checkmark & \xmark & \xmark & \checkmark  & \checkmark & 0 & Role-play Interaction \\

\midrule
\textbf{CEAEval (ours)}
& \checkmark & \checkmark & \checkmark & \checkmark  & \checkmark & 15 & \makecell{Contextual Expressive\\Appropriateness} \\
\bottomrule
\end{tabular}
\vspace{-0.2cm}
\caption{Comparison of CEAEval with existing expressive speech evaluation benchmarks. Long-range context refers to conversational contexts exceeding 10 dialogue turns.}
\vspace{-0.2cm}
\label{tab:context_eval_comparison}
\end{table*}

Traditional speech evaluation methods primarily focus on word accuracy \cite{seedtts}, naturalness \cite{qualitynet}, signal quality \cite{reddy2021dnsmos}, or emotional intensity \cite{im2022emoq} at the utterance level. 
However, these metrics are insufficient for determining whether speech expressiveness aligns with contextual intent.
Expressive appropriateness can only be meaningfully assessed once conversational context and discourse progression are made explicit, as these factors strongly constrain the range of appropriate expressive realizations \cite{tawari2010speech}.
As illustrated in Figure~\ref{fig:context-example}, Joy’s utterance could be perceived as reproachful or even angry when considered in isolation; however, within the given conversational context, an expressive realization conveying restrained amusement is the most appropriate, which cannot be captured by existing evaluation methods based on emotion classification or intensity prediction \cite{ma2024emotion2vec, zhou2022emotion}.

Despite increasing interest in expressive speech evaluation, existing resources remain insufficient for studying expressive appropriateness under rich contextual settings. As summarized in Table~\ref{tab:context_eval_comparison}, prior benchmarks primarily target speech naturalness, dialogue quality, or role-play interaction, and vary substantially in their use of generated speech or context, long-range discourse, and annotation granularity.

From a data perspective, most existing datasets either focus on daily conversational speech with limited expressive range \cite{yan2025uro} or rely on synthesized speech \cite{ji2025wavreward} and generated contexts to approximate context–expression alignment \cite{zhan2025vstyle, jiang2025speechrole}. As a result, expressive behavior is often evaluated without grounding in authentic narrative structure or long-range discourse, which are crucial for determining whether a given expressive realization is contextually appropriate.

From a methodological perspective, most existing speech evaluation approaches operate at the single-utterance or short-context level, even when limited multi-turn dialogue is considered \cite{zhang2025speechjudge}. Contextual information is often summarized or truncated, which hinders modeling long-range dependencies between speech expressiveness and narrative progression \cite{speechdrame}.
Recent work has begun to incorporate large language models (LLM) and chain-of-thought (CoT) reasoning to enable semantic- and discourse-level analysis \cite{ji2025wavreward, zhang2025speechjudge}. However, particularly for base models with limited reasoning capacity, directly applying long textual reasoning to speech evaluation can cause attention to be dominated by text, suppressing speech perception under long-context multimodal inputs \cite{tian2025stepaudior1technicalreport}. This fundamentally constrains effective expressive appropriateness evaluation, which requires joint reasoning over both speech and long-range context.

To address these challenges, we introduce CEAEval, a unified framework for context-rich expressive appropriateness evaluation of Mandarin speech under rich contextual settings. CEAEval integrates long-range context modeling, fine-grained expressive perception, and stable reasoning.
Our main contributions are threefold:
\begin{itemize}[itemsep=0.0em, topsep=0pt]
\item We formalize context-rich expressive appropriateness evaluation and introduce the first human-annotated dataset for this task, based on real Mandarin audiobook speech. The dataset contains long-range narrative context and 15 carefully designed annotation dimensions with high inter-annotator consistency.
\item We propose a planner–judge decoupled evaluation framework for expressive appropriateness, which separates long-context textual reasoning from fine-grained perceptual scoring of speech. To alleviate text-dominated reasoning under long-context inputs, we further introduce an adaptive audio attention bias mechanism and reinforcement learning optimization.
\item Experiments on the proposed task demonstrate that our method maintains strong agreement with human judgments as contextual length increases, achieving a linear correlation coefficient of 0.72 and an accuracy of 70.8\%, while providing interpretable scoring rationales. Our demo, annotated data, model, and code will be released at
\url{https://wangtianrui.github.io/ceaeval/}.
\end{itemize}

\section{Related Work}

\subsection{General Speech Evaluation}
A large body of prior work on speech evaluation focuses on intelligibility~\cite{maniati2022somos}, naturalness~\cite{mittag2021nisqa}, perceived quality~\cite{wang2025qualispeech}, and robustness to noise~\cite{reddy2021dnsmos}, typically supported by human perceptual ratings. While effective for assessing acoustic quality and signal-level properties, these approaches are not designed to evaluate whether a speech sample is expressively appropriate for its contextual setting, as they largely operate at the utterance level and do not incorporate rich discourse or narrative context.

\subsection{Expressive Speech Evaluation and Data}
Recent work has begun to explore expressive aspects of speech, including aesthetics and prosody~\cite{yao2025songeval}, emotional cues in dialogue~\cite{yan2025uro, ji2025wavreward}, and human preference judgments~\cite{zhang2025speechjudge}. Some studies further incorporate role- or scene-related information to enrich expressive modeling~\cite{jiang2025speechrole, speechdrame}. 
Despite these advances, existing expressive speech datasets remain insufficient for context-rich expressive appropriateness evaluation. As summarized in Table~\ref{tab:context_eval_comparison}, fine-grained human annotations under real speech, long-range contextual settings are largely absent.

\subsection{Learning-based Speech Evaluation with Large Language Models}
Recent speech evaluation methods increasingly adopt learning-based approaches to predict human perceptual judgments. Models such as Quality-Net~\cite{qualitynet}, DNSMOS~\cite{reddy2021dnsmos}, and MOSNet~\cite{lo2019mosnet} primarily focus on acoustic quality and signal-level attributes. To incorporate semantic- and discourse-level information, several recent works integrate large language models and CoT-style reasoning into speech evaluation frameworks~\cite{ji2025wavreward, zhang2025speechjudge, wang2025qualispeech, jiang2025speechrole, speechdrame, manku2025emergenttts}. However, these approaches typically rely on fine-tuning LLMs on speech modalities, which has been shown to degrade their original text reasoning capabilities~\cite{tang2024salmonn}. This limitation poses a fundamental challenge for expressive appropriateness evaluation, which requires robust reasoning over rich textual context and discourse structure.

\begin{figure*}[t]
  \centering
  \includegraphics[width=0.99\textwidth]{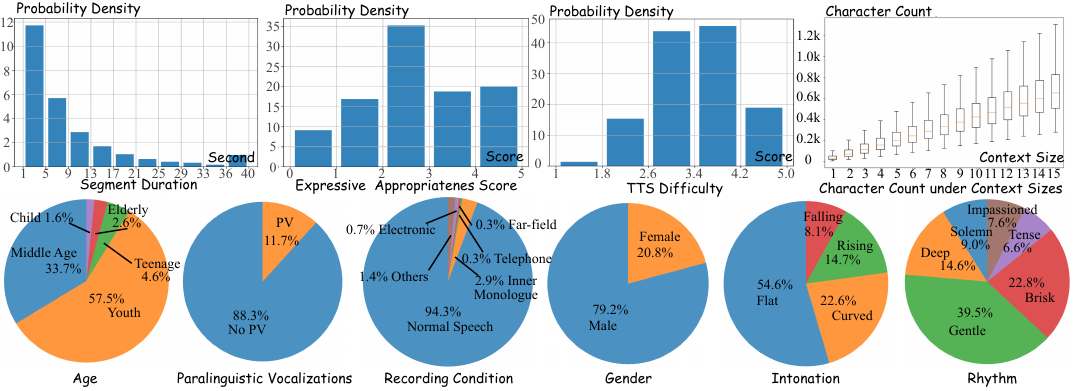}
  \vspace{-0.2cm}
  \caption{Statistical distribution of annotation categories and attributes in the CEAEval-D dataset.}
  \label{fig:datashow}
\end{figure*}
\section{Proposed Method}
\subsection{Task Definition and Problem Formulation}
\label{pf}

We study the task of context-rich expressive appropriateness evaluation for Mandarin conversational speech, which aims to assess whether the expressive realization of a spoken utterance aligns with the underlying content, discourse intent, and situational context given its rich narrative and conversational background.
Following established principles in Chinese broadcast speech and reading aesthetics \cite{zhang2003chinese}, we assess expressive appropriateness by jointly considering emotional expression, prosodic realization (intonation and rhythm), recording conditions, and the appropriateness of paralinguistic vocalizations. These factors are evaluated in an integrated manner rather than in isolation, reflecting how expressiveness is perceived in natural speech.
Detailed definitions and analyses of how each expressive attribute contributes to overall expressive appropriateness are provided in Appendix~\ref{appendixdataanno}.
The evaluation output consists of a scalar appropriateness score ranging from 0 to 5, along with a structured reasoning process that analyzes relevant paralinguistic and expressive cues. 
Such rationales are intended to provide interpretable references for downstream expressive speech evaluation and generation tasks.

\subsection{CEAEval-D: Dataset}
Evaluating expressive appropriateness under rich contextual settings requires speech data that jointly provide authentic expressive realizations, long-range narrative context, and reliable human judgments. To support this task, we construct CEAEval-D based on narrated Mandarin audiobooks, which naturally exhibit rich discourse structure, diverse speaker roles, and context-dependent expressive variation.
Specifically, we collect 84 audiobook works (including 2 high-quality TTS-generated works), resulting in a total of 3,505 hours of performed speech. From this corpus, we strictly select speech segments that are publicly accessible and suitable for release to construct a subset for manual annotation. These annotated segments are drawn from contiguous portions of each work and are accompanied by complete story texts, enabling reliable context construction and expressive assessment. All manually annotated data are curated in accordance with ethical research and data privacy considerations. The data selected for manual annotation comprise 16.1 hours of speech, split into 14.65 hours for training and 1.45 hours for evaluation, with zero overlap in speech samples.

\subsubsection{Weak Annotation}
\label{weakannotation}
To enable context-rich expressive appropriateness evaluation at scale, we generate weak descriptive annotations for the full 3,505-hour corpus using Qwen3-Omni-Captioner~\cite{ma2025omnicaptionerdatapipelinemodels}. These captions provide detail captions of speech and are used to distill descriptive and reasoning capabilities into the judge model.  

Prior to manual annotation, we apply an automatic speech recognition (ASR) model \cite{gao23g_interspeech} to pre-segment the 16.1 hours of selected audio and generate preliminary content annotations, which serve as reference material to facilitate and standardize subsequent human annotation.
\subsubsection{Manual Annotation}
\label{manualannotation}
The selected 16.1 hours of speech data are manually annotated to provide reliable supervision for context-rich expressive appropriateness evaluation. Annotation is conducted by 18 native Mandarin-speaking graduate students with backgrounds in speech emotion research, following unified annotation guidelines and a standardized calibration protocol, with inter-annotator agreement verified on a shared calibration subset (see Appendix~\ref{appendixdataanno}).

Within each selected excerpt, speech is further segmented into fine-grained utterances. Each utterance is annotated with a multidimensional set of attributes (detailed in Appendix~\ref{appendixdataanno}), including expressive appropriateness scores, intonation, rhythm, emotion categories, refined textual context, TTS difficulty, recording conditions, background music presence, paralinguistic vocalizations, and sound events. In addition, auxiliary information such as utterance boundaries, refined textual content, and speaker metadata (role name, gender, and age) is also provided. Together, these annotations capture complementary aspects of expressive behavior relevant to appropriateness judgments under rich contextual settings.

As illustrated in Figure~\ref{fig:datashow}, we present summary statistics of key annotation dimensions and contextual properties in the dataset, which spans a wide range of context sizes, prosodic patterns, and expressive conditions and thus enables evaluation under diverse discourse settings. In this work, context size (CTS) denotes the number of consecutive dialogue or narrative lines provided as contextual input for a target utterance (see Appendix~\ref{app:context}), with CTS=0 corresponding to the context-free setting.

\begin{figure*}[t]
  \centering
  \includegraphics[width=1.0\textwidth]{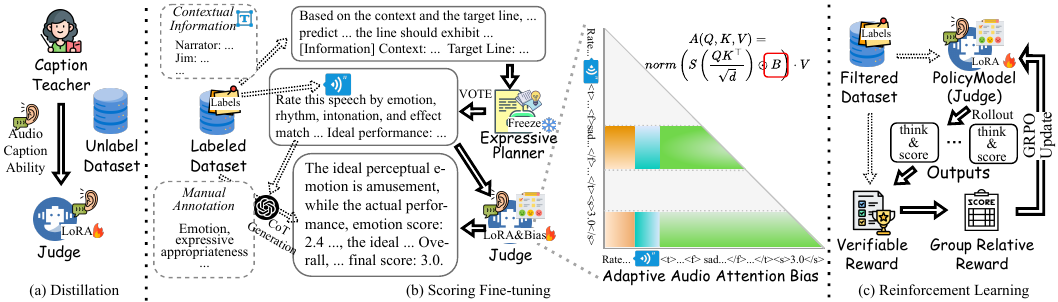}
  \vspace{-0.7cm}
  \caption{Overview of CEAEval-M, which is trained through a three-stage pipeline for context-rich expressive appropriateness evaluation. Dashed arrows indicate data flow, while solid arrows denote inference or training flow.}
  \vspace{-0.2cm}
  \label{fig:main}
\end{figure*}

\subsection{CEAEval-M: Speech-LLM as a Judge}

We propose CEAEval-M, a speech-LLM that evaluates expressive appropriateness by jointly reasoning over speech signals and rich textual context. As shown in Figure~\ref{fig:main}, the model is trained through a three-stage pipeline.
First, we distill audio perceptual reasoning abilities from a captioning teacher using 3505-hours data, enabling the model to recognize expressive cues and paralinguistic events in speech. Next, a frozen text-only expressive planner predicts an ideal expressive profile implied by the contextual text, which serves as a reference for appropriate expression. The Speech-LLM judge is then fine-tuned via LoRA to compare the observed speech realization with this planned expressiveness and to produce an appropriateness score in a CoT style, supported by a learnable audio attention bias. Finally, we apply reinforcement learning to further improve scoring robustness and calibration with respect to human judgments.

\subsubsection{Expressive Planner}
The expressive appropriateness scoring task requires joint reasoning over a speech segment and its narrative context, which may span long sequences of text. As shown in Figure~\ref{fig:datashow}, when the context size reaches $15$, the accumulated narrative context can exceed 1,200 characters, posing challenges for robust discourse-level modeling. Given the limited amount of textual knowledge available from annotated speech in our 16.1-hour corpus, such supervision is insufficient to bridge the gap between Omni-style multimodal models and text-only LLMs in long-range textual modeling.
We therefore introduce a dedicated text-only large language model, Qwen3-8B \cite{yang2025qwen3}, as an expressive planner to explicitly model narrative context and infer ideal speech expressiveness. The planner takes the narrative context and target utterance as input and predicts an ideal expressive profile covering emotion, rhythm, intonation, and recording condition. To improve robustness under varying context sizes, we construct cumulative context windows ranging from one to fifteen preceding lines and aggregate the resulting predictions through a voting strategy, which mitigates instability caused by context truncation or local ambiguity. Details of the voting procedure are provided in Appendix~\ref{appendixplanner}.

\subsubsection{Knowledge Distillation}
\label{kd}
To focus the model on expressive speech perception, we use Qwen3-Omni-Captioner to generate weak descriptive annotations for 3,505 hours of speech data and perform distillation-based training with Qwen2.5-Omni-7B \cite{xu2025qwen2} equipped with LoRA \cite{hu2022lora}, as shown in Figure~\ref{fig:main}.

\subsubsection{Scoring Model with CoT Supervision}
\label{cottrain}

Building on the distilled backbone, we train a judge model for expressive appropriateness evaluation under rich contextual conditions. The judge conditions on planner outputs together with the input speech and performs expressiveness analysis in a CoT manner, covering emotion, recording conditions, rhythm, intonation, paralinguistic vocalizations, and sound events (see Appendix~\ref{app:planprompt}).
To supervise CoT reasoning, we generate CoT annotations using GPT-4o based on ground-truth scores, manually annotated expressive attributes, and the outputs of the expressive planner (details in Appendix~\ref{app:cot}). The judge model is then fine-tuned with LoRA on the resulting dataset to learn structured reasoning for expressive appropriateness evaluation.

\subsubsection{Adaptive Audio Attention Bias}
\label{bias}
While CoT supervision improves reasoning transparency, it significantly increases textual length. For base models with limited reasoning capacity, this often induces text-dominated shortcut reasoning, causing the model to under-attend to speech signals \cite{wang2025pay, sim2025can, tian2025stepaudior1technicalreport}. To counter this effect, we introduce an adaptive audio attention bias into the self-attention computation of the Speech-LLM, as illustrated in Figure~\ref{fig:main}. The modified attention operation is defined as:
\begin{equation}
\label{eqbias}
\begin{aligned}
\mathrm{A}(Q, K, V)
&=
\mathrm{norm}
\Bigl(
\mathrm{S}
\Bigl(
\frac{QK^{\top}}{\sqrt{d}}
\Bigr)
\odot B
\Bigr)
V ,
\end{aligned}
\end{equation}
where \( \mathrm{A}(\cdot) \) denotes the attention operation, 
\( \mathrm{S}(\cdot) \) is the softmax function,
\( Q \), \( K \), and \( V \) denote the query, key, and value matrices,
\( d \) denotes the hidden dimension,
\( \odot \) denotes element wise multiplication,
and \( B \) denotes an adaptive attention bias:
\begin{equation}
\begin{aligned}
B
&=
2 \cdot \sigma\!\bigl(f_{\mathrm{p}}(X)\bigr) \cdot M_{\mathrm{p}} \\
&\quad
+
\bigl(1 + \sigma\!\bigl(f_{\mathrm{a}}(X)\bigr)\bigr) \cdot M_{\mathrm{a}} \\
&\quad
+
\sigma\!\bigl(f_{\mathrm{CoT}}(X)\bigr) \cdot M_{\mathrm{CoT}} 
+
M_{\mathrm{base}} ,
\end{aligned}
\end{equation}
where \( X \) denotes the input hidden state.
\( f_{\mathrm{p}} \), \( f_{\mathrm{a}} \), and \( f_{\mathrm{CoT}} \) denote learnable linear projections that map the input feature to a scalar, 
and \( \sigma(\cdot) \) denotes the sigmoid function.
The four binary masks \( M_{\mathrm{p}} \), \( M_{\mathrm{a}} \), \( M_{\mathrm{CoT}} \), and \( M_{\mathrm{base}} \) respectively indicate system prompt regions, audio regions that require focused attention, CoT regions, and remaining regions that remain unchanged.
By adapting attention weights through region-specific bias, the proposed model mitigates audio attention dilution caused by increased textual inputs and improves score prediction accuracy under CoT-style supervision. Implementation details of the attention bias construction and mask definitions are provided in Appendix~\ref{appendixbias}.

\subsubsection{Reinforcement Learning Optimization}

Our final objective is accurate and stable expressive appropriateness score prediction. While CoT-supervised training yields reasonable behavior, classification-style supervision does not explicitly model distances between continuous scores, leading to instability. We therefore optimize the judge model using GRPO~\cite{guo2025deepseek} on a filtered and balanced train-set to directly optimize distance-aware score prediction, as shown in Figure~\ref{fig:main}. Details of the filtering and resampling strategy are provided in Appendix~\ref{appendix:filter}.
We define a reward function that combines regression accuracy \cite{ji2025wavreward} and bucket-level ordinal consistency:
\begin{equation}
\begin{aligned}
r(\hat{s}, s)
&=
\exp\!\Bigl(
-\frac{|\hat{s} - s|}{\sigma}
\Bigr)
+
\exp\!\bigl(
-|b(\hat{s}) - b(s)|
\bigr),
\end{aligned}
\end{equation}
where \( \hat{s} \) and \( s \) denote the predicted and ground truth scores, wrapped with \texttt{<s>} and \texttt{</s>} in the output sequence, \( \sigma = 1.0 \), and \( b(\cdot) \) maps scores to discrete buckets:
\begin{equation}
b(s) = \min\!\bigl(5,\;\max\!\bigl(0,\;\lfloor s \rfloor\bigr)\bigr).
\end{equation}
The GRPO objective is defined as:
\begin{equation}
\max_{\theta}
\;
\mathbb{E}_{y \sim \pi_\theta}
\Bigl[
\mathrm{clip}(r, -\epsilon, \epsilon)
-
\beta \,
\mathrm{KL}
\bigl(
\pi_\theta \,\|\, \pi_{\mathrm{ref}}
\bigr)
\Bigr],
\end{equation}
where \( \epsilon = 0.1 \) and \( \beta = 0.01 \).
The reference policy \( \pi_{\mathrm{ref}} \) shares the same backbone as \( \pi_\theta \) but excludes LoRA parameters.

\begin{table*}[t]
\centering
\small
\setlength{\tabcolsep}{2.5pt}
\resizebox{\textwidth}{!}{
\begin{tabular}{l|cccccccc|cccccccc}
\toprule
\multirow{4}{*}{Model}
& \multicolumn{8}{c|}{w/o CoT}
& \multicolumn{8}{c}{w CoT} \\
\cmidrule(lr){2-9} \cmidrule(lr){10-17}

& \multicolumn{4}{c}{LCC $\uparrow$}
& \multicolumn{4}{c|}{ACC \% $\uparrow$}
& \multicolumn{4}{c}{LCC $\uparrow$}
& \multicolumn{4}{c}{ACC \% $\uparrow$} \\
\cmidrule(lr){2-5} \cmidrule(lr){6-9}
\cmidrule(lr){10-13} \cmidrule(lr){14-17}

\makebox[0.12\linewidth][r]{CTS}
& 0 & 5 & 10 & 15
& 0 & 5 & 10 & 15
& 0 & 5 & 10 & 15
& 0 & 5 & 10 & 15 \\
\midrule

Qwen2.5-Omni
& 0.09 & 0.16 & 0.15 & 0.07
& 28.01 & 27.85 & 27.69 & 28.99
& 0.03 & 0.14 & 0.11 & 0.09
& 28.11 & 31.43 & 30.94 & 28.83 \\

Kimi-Audio
& -0.01 & 0.17 & 0.19 & 0.13
& \textbf{35.67} & 36.32 & 33.71 & 31.60
& 0.04 & 0.06 & 0.04 & 0.03
& 30.13 & 29.48 & 26.22 & 26.71 \\

Phi-4-MM
& -0.01 & 0.27 & 0.18 & 0.10
& 33.39 & 33.22 & 25.20 & 32.08
& 0.00 & 0.07 & 0.06 & 0.06
& 32.74 & \textbf{34.69} & 30.78 & 32.90 \\

Gemma-3n
& 0.15 & 0.21 & 0.10 & 0.06
& 30.78 & \textbf{40.07} & 29.15 & 28.34
& 0.14 & 0.10 & 0.04 & 0.07
& 31.27 & 34.04 & \textbf{34.85} & 32.84 \\

Step-Audio-R1
& 0.16 & 0.12 & 0.16 & 0.11
& 28.83 & 27.69 & 26.22 & 27.20
& 0.08 & 0.17 & 0.03 & 0.07
& 22.80 & 25.73 & 24.27 & 24.43 \\

Midashenglm
& 0.09 & 0.24 & 0.16 & 0.17
& 22.80 & 28.99 & 32.90 & 29.80
& -0.04 & 0.19 & 0.19 & 0.09
& 21.82 & 28.83 & 34.04 & \textbf{35.71} \\

GPT-4o-Audio
& 0.08 & 0.09 & 0.13 & 0.06
& 29.99 & 31.82 & 31.66 & 32.73
& 0.11 & 0.18 & 0.19 & \textbf{0.22}
& 23.16 & 22.76 & 26.71 & 27.41 \\

Gemini-3-Pro
& 0.11 & 0.14 & 0.16 & 0.10
& 29.41 & 27.69 & 26.23 & 26.41
& 0.09 & 0.17 & 0.14 & 0.12
& 22.41 & 20.43 & 18.73 & 21.17 \\

Voxtral-Mini
& 0.20 & \textbf{0.33} & 0.23 & 0.22
& 32.74 & 31.11 & 31.27 & 30.94
& 0.04 & 0.02 & 0.15 & 0.06
& \textbf{33.39} & 32.57 & 31.76 & 29.80 \\

Qwen3-Omni
& \textbf{0.21} & 0.27 & \textbf{0.25} & \textbf{0.28}
& 35.34 & 34.85 & \textbf{35.18} & \textbf{33.55}
& \textbf{0.21} & \textbf{0.30} & \textbf{0.29} & \textbf{0.22}
& 24.42 & 29.49 & 32.74 & 30.13 \\

\midrule
 CEAEval-M
& \textbf{0.54} & \textbf{0.58} & \textbf{0.61} & \textbf{0.61}
& \textbf{59.96} & \textbf{62.00} & \textbf{64.11} & \textbf{65.47}
& \textbf{0.61} & \textbf{0.69} & \textbf{0.71} & \textbf{0.72}
& \textbf{64.33} & \textbf{68.47} & \textbf{70.12} & \textbf{70.80} \\
\bottomrule
\end{tabular}
}
\vspace{-0.3cm}
\caption{Performance comparison on contextual speech expressiveness appropriateness assessment, evaluated with and without CoT across different context sizes (CTS).}
\vspace{-0.2cm}
\label{tab:cot_full}
\end{table*}

\section{Experiment Setup}

\subsection{Data}
As described in Section~\ref{kd}, the distillation stage uses 3,505 hours of unlabeled audiobook speech. For context-rich expressive appropriateness scoring, we annotate 16.1 hours of speech, split into 14.65 hours for training and 1.45 hours for testing, with strict story-level separation and no overlap in narratives, characters, or scenes.

\subsection{Models and Training Settings}
We adopt Qwen3-8B as the expressive planner and Qwen2.5-Omni-7B-Thinker as the backbone of the judge model. All fine-tuning and reinforcement learning stages use LoRA, with the rank set to 32 and the scaling factor alpha set to 64. The learning rate linearly increases to a peak value of $5\times10^{-6}$ during the first 10\% of training steps and then gradually decays to $5\times10^{-7}$ by the end of training. Training runs on eight NVIDIA A40 GPUs, with a per-GPU batch size of 4.
To support multilingual instructions and outputs, we design language-specific system prompts for Chinese and English, with details provided in the Appendix~\ref{app:planprompt}.

\subsection{Baselines and Metrics}
Since existing public evaluation models don't support rich narrative context as defined in our task, and our approach adopts commonly used strategies such as supervised fine-tuning, CoT reasoning, and reinforcement learning, we focus our comparison on representative speech-capable models.
Specifically, we evaluate Gemma-3n \cite{team2024gemma}, Midashenglm \cite{dinkel2025Midashenglm}, Phi-4-MM \cite{abdin2024phi}, Qwen2.5-Omni-7B \cite{xu2025qwen2}, Qwen3-Omni-30B-Instruct \cite{yang2025qwen3}, Voxtral-Mini \cite{liu2025voxtral}, Step-Audio-R1 \cite{tian2025stepaudior1technicalreport}, Kimi-Audio \cite{kimiteam2025kimiaudiotechnicalreport}, GPT-4o-Audio \cite{hurst2024gpt}, and Gemini-3-Pro \cite{comanici2025gemini25pushingfrontier}.

We evaluate score prediction using the Linear Correlation Coefficient (LCC) \cite{pearson} and a tolerance-based accuracy metric (ACC) \cite{itu_p800}, where a prediction is considered correct if the absolute difference between the predicted score and the annotated score is within 1. 
LCC serves as the primary metric for assessing score consistency, while ACC provides a complementary measure of absolute prediction error.

\section{Results}
\subsection{Context-rich Speech Expressiveness Appropriateness Evaluation}
\label{sec:experiments}
We evaluate different models on contextual speech expressiveness appropriateness assessment under varying context sizes (CTS), as shown in Table~\ref{tab:cot_full}.
To account for prompt language sensitivity, we test both Chinese and English prompts and report the best-performing language for each model.
We further compare direct scoring with CoT reasoning and analyze performance trends as CTS increases.
Details of context construction and prompt designs are provided in Appendix~\ref{app:context} and \ref{app:prompts}, and a detailed comparison of model parameter counts is provided in Appendix~\ref{app:params}.

Across all evaluated models, the context-free baseline (CTS=0) consistently underperforms settings with contextual input. As CTS increases from 0 to moderate values, most models show clear improvements in both LCC and ACC, indicating that narrative context plays a critical role in aligning speech expressiveness with communicative intent. This trend underscores the importance of contextual information for expressive appropriateness evaluation, even when existing models cannot fully exploit long-range context.

As CTS increases, most baseline models exhibit a rise-then-decline performance pattern.
Moderate context initially improves alignment between speech expressiveness and narrative intent, whereas longer contexts degrade performance.
As illustrated in Fig.~\ref{fig:datashow}, when CTS exceeds 5, the contextual text rapidly grows beyond 300 characters, causing multimodal models to become increasingly text-dominated and less sensitive to acoustic cues, which degrades evaluation performance \cite{wang2025pay, liu2024paying, tian2025stepaudior1technicalreport}.
In contrast, our method achieves consistently higher and more stable performance across all context sizes, reaching an ACC of 70.80\% and an LCC of 0.72.
By decoupling context modeling from speech scoring via an expressive planner and explicitly rebalancing audio attention in the judge model, our approach enables CoT-style reasoning without overwhelming the speech modality.

\begin{table}[t]
\vspace{0.2cm}
\centering
\small
\setlength{\tabcolsep}{4.5pt}
\begin{tabular}{lcccc}
\toprule
Model
& \multicolumn{2}{c}{w/o CoT}
& \multicolumn{2}{c}{w CoT} \\
\cmidrule(lr){2-3} \cmidrule(lr){4-5}

& LCC $\uparrow$ & ACC \% $\uparrow$
& LCC $\uparrow$ & ACC \% $\uparrow$ \\
\midrule

Gemma-3n
& 0.21 & 48.70
& 0.15 & 32.74 \\

Kimi-Audio
& 0.20 & 48.21
& 0.09 & 42.35 \\

Phi-4-MM
& 0.26 & 46.58
& 0.15 & 46.25 \\

Midashenglm
& 0.27 & 44.30
& 0.19 & 44.46 \\

Step-Audio-R1
& 0.21 & 40.72
& 0.16 & 42.18 \\

Qwen2.5-Omni
& 0.24 & 49.89
& 0.27 & 42.35 \\

Voxtral-Mini
& 0.32 & 54.37
& 0.24 & 45.86 \\

GPT-4o-Audio
& 0.25 & 47.04
& 0.29 & 44.95 \\

Gemini-3-Pro
& 0.22 & 51.21
& 0.25 & 48.81 \\

Qwen3-Omni
& \textbf{0.36} & \textbf{58.17}
& \textbf{0.30} & \textbf{49.47} \\

\midrule
 CEAEval-M
& \textbf{0.61} & \textbf{65.47}
& \textbf{0.72} & \textbf{70.80} \\
\bottomrule
\end{tabular}
\vspace{-0.2cm}
\caption{Planner-assisted evaluation with and without CoT.}
\vspace{-0.2cm}
\label{tab:cot_summary}
\end{table}

\subsection{Evaluation Baselines with Planner}

To isolate the effect of the expressive planner, we evaluate all baseline models under the planner-assisted setting, as shown in Table~\ref{tab:cot_summary}. Compared to direct contextual conditioning in Table~\ref{tab:cot_full}, planner-assisted evaluation consistently improves performance and reduces variance across models, indicating that abstracting narrative context into structured expressive plans leads to more reliable expressive appropriateness evaluation. This effect becomes particularly evident under long-context settings: when the number of context segments reaches 15, raw contextual text often exceeds 600 characters, a regime in which baseline models tend to exhibit unstable or text-dominated reasoning. By converting long narrative inputs into semantically grounded plans through multi-context voting, the expressive planner alleviates the burden of long-text processing in scoring models. Notably, the planner operates as a text-only module, allowing speech-centric scoring models to focus on acoustic perception while still benefiting from rich contextual information.

\subsection{Ablation Study}
\begin{figure}[t]
  \centering
  \includegraphics[width=0.48\textwidth]{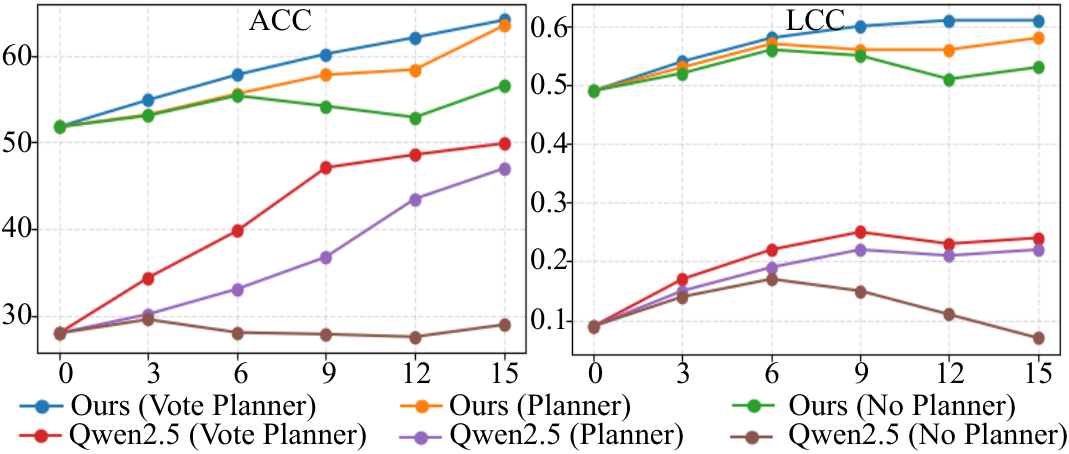}
  \vspace{-0.6cm}
  \caption{Performance trends under increasing context size.}
  \label{fig:voting}
\end{figure}

We conduct ablation experiments to analyze the contribution of each proposed component, with results summarized in Table~\ref{tab:ablation} and Figure~\ref{fig:voting}.
Each model configuration is indexed by its ID in the table.
For configurations involving CoT supervision, inference is also performed in a CoT manner to ensure consistency between training and evaluation.

\textbf{Knowledge distillation.}
We first examine the effect of knowledge distillation.
Comparing ID~(1) and ID~(2), distilling expressive perception knowledge from Qwen3-Omni into Qwen2.5-Omni leads to a clear improvement in both LCC and ACC.
This result indicates that initializing the scoring model with stronger audio caption capability is beneficial for expressive appropriateness evaluation.
Notably, Qwen3-Omni already demonstrates competitive performance in our task (Table~\ref{tab:cot_full}), motivating its use as the teacher model for distillation.

\begin{table}[t]
\centering
\small
\resizebox{0.48\textwidth}{!}{
\setlength{\tabcolsep}{1pt}
\begin{tabular}{ccccccc|cc}
\toprule
ID & Distill. & CoT & Planner & AttenBias & RL &  & LCC $\uparrow$ & ACC \% $\uparrow$ \\
\midrule

(0) & \multicolumn{5}{c}{Baseline-Qwen2.5-Omni(w/o-SFT)} &  & 0.09 & 28.83 \\
\midrule

(1) & No  & No      & No      & No  & No  &  & 0.45 & 48.49 \\
(2) & Yes & No      & No      & No  & No  &  & 0.53 & 56.55 \\
(3) & Yes & No      & Only15  & No  & No  &  & 0.58 & 63.47 \\
(4) & Yes & No      & VOTE    & No  & No  &  & 0.61 & 64.11 \\
(5) & Yes & No      & GPT4o    & No  & No &  & 0.63 & 65.03 \\
(6) & Yes & No      & VOTE    & No  & Yes &  & 0.65 & 66.86 \\
\midrule

(7) & Yes & Yes     & VOTE    & No  & No  &  & 0.40 & 49.09 \\
(8) & Yes & Yes+No  & VOTE    & No  & No  &  & 0.41 & 50.17 \\
(9) & Yes & Yes+No  & VOTE    & No  & Yes &  & 0.47 & 54.44 \\
\midrule

(10)  & Yes & Yes     & VOTE   & Yes & No  &  & 0.61 & 64.07 \\
(11) & Yes & Yes+No  & VOTE   & Yes & No  &  & 0.64 & 67.33 \\
(12) & Yes & Yes+No  & VOTE   & Yes & Yes &  & \textbf{0.72} & \textbf{70.80} \\
\bottomrule
\end{tabular}
}
\vspace{-0.2cm}
\caption{Ablation study on the effects of distillation, chain-of-thought (CoT), planner, attention bias, and reinforcement learning (RL).}
\label{tab:ablation}
\end{table}

\textbf{Expressive planner and context size.}
We analyze the effect of the expressive planner under increasing context length in Figure~\ref{fig:voting}. Without the planner, both Qwen2.5-Omni and our model exhibit unstable performance as context size grows, following a rise-then-decline pattern (green and brown lines). 
Introducing the expressive planner fundamentally changes this behavior: with planner-based abstraction, performance increases more steadily and gradually converges. This trend is further supported by the quantitative comparison between ID~(2) and ID~(3) in Table~\ref{tab:ablation}, indicating that summarizing long narrative context into structured expressiveness plans reduces the burden of long-context reasoning in the speech-LLM model. 
To further mitigate variability across context configurations, we introduce a voting mechanism. Comparing ID~(3) and ID~(4), voting consistently improves stability and overall performance, which is also reflected by smoother trends in Figure~\ref{fig:voting}.

We additionally evaluate GPT-4o as an alternative expressive planner without voting (ID~(5)). Although it yields higher performance, the Qwen3-8B-based voting planner achieves comparable results. We therefore adopt the voting-based planner as the default setting, balancing performance, reproducibility, and deployment cost.

\textbf{Reinforcement learning.}
We further introduce reinforcement learning to directly optimize score prediction accuracy.
Comparisons between ID~(4) and ID~(6), ID~(8) and ID~(9), as well as ID~(11) and ID~(12), show consistent performance gains from reinforcement learning.
These results indicate that distance-aware optimization further improves numerical stability on top of supervised fine-tuning.

\textbf{CoT and attention bias.}
Finally, we examine the interaction between CoT-style supervision and audio attention.
As shown by ID~(7), ID~(8), and ID~(9), introducing CoT-style reasoning alone leads to noticeable performance degradation from ID~(4).
This is caused by the increased amount of textual content produced during CoT reasoning, which, given the limited reasoning capacity of the base model (Qwen2.5-Omni), shifts its attention away from the speech modality.
To address this issue, we introduce an adaptive attention bias mechanism.
Comparisons between ID~(7) and ID~(10), as well as between ID~(8) and ID~(11), show that attention bias effectively counteracts the negative impact of CoT supervision and restores performance.
Moreover, when CoT and non-CoT supervision are jointly applied during training (ID~(11)), the resulting model outperforms its non-CoT counterpart (ID~(4)), indicating that CoT provides complementary benefits when its modality imbalance is properly controlled.
The introduction of the expressive planner and CoT reasoning not only improves scoring accuracy but also provides high interpretability. A concrete example of this reasoning process is presented in Appendix \ref{app:case_study}, which compares the ideal expressive plan with the actual speech realization.

\section{Conclusion}

This paper introduces expressive appropriateness evaluation for Mandarin speech under rich narrative context and presents the first systematic study of this problem from the perspectives of task formulation, data construction, and model design. We define expressive appropriateness as the alignment between speech realization and the latent communicative intent implied by contextual narratives, rather than isolated acoustic attributes.
To support this task, we construct a real-speech dataset comprising 16.1 hours of contextualized audiobook speech, annotated with 15 carefully designed dimensions and exhibiting high inter-annotator consistency. Building on this dataset, we propose a context-rich evaluation framework that integrates knowledge distillation, planner–scorer decoupling, adaptive audio attention bias, and reinforcement learning. This design enables robust reasoning over long-range narrative context while preserving sensitivity to speech signals, resulting in the predicted expressive appropriateness scores that closely align with human judgments.
Beyond audiobook, the proposed framework provides a general diagnostic tool for expressive speech generation in dialogue systems, offering a principled way to assess whether generated speech appropriately reflects contextual intent.

\section*{Limitations}
While this work presents a comprehensive framework for context-rich expressive appropriateness evaluation, several limitations remain. 
First, as expressive appropriateness is shaped by language-specific and cultural factors, our current study focuses on Mandarin speech. In future work, we plan to extend the proposed framework to additional languages and cultural contexts and scale up the annotated dataset size, with appropriate adaptation to account for cross-linguistic and cross-cultural variation in expressive intent.
Second, we primarily model contextual information from textual narratives, and incorporating speech-level context across neighboring utterances may further enhance expressive evaluation. 
Finally, although human annotation improves reliability, expressive appropriateness remains inherently subjective, and automatic scores should be interpreted with caution rather than used as the sole criterion for real-world decision-making.

\section*{Ethical Statement}

\paragraph{Human annotation and fair compensation.}
All human annotators involved in data creation were native Mandarin speakers with academic backgrounds in speech emotion or affective speech research.
Annotators were either legally employed graduate students supported by formal scholarships, and all participated as co-authors of this work.
Annotation work was compensated in accordance with local minimum wage regulations and institutional guidelines.
This compensation scheme aligns with ACL requirements regarding fair treatment and remuneration of human participants.

\paragraph{Data privacy and consent.}
We will not release the full 3,505 hours of speech data.
Only the manually annotated subset will be made available.
All annotated speech segments are carefully reviewed and are derived from publicly accessible, user-uploaded audio content on platforms such as Bilibili.
Each released audiobook segment will be shorter than 10 minutes, which is substantially shorter than preview excerpts commonly provided by commercial audiobook platforms, thereby minimizing potential copyright and privacy risks.
The released data will not contain any personally identifiable information or sensitive user data.

\paragraph{Licensing and responsible use.}
The manually annotated dataset subset will be released under a CC-BY-NC license.
This license explicitly restricts usage to non-commercial academic research and is consistent with ACL guidelines on ethical dataset release and respect for copyright.
Users of the dataset are required to adhere to the license terms and applicable regulations.

\paragraph{Model release for reproducibility.}
While the full 3,505 hours of speech data will not be publicly released, the distilled model checkpoints and final model parameters trained on this data will be made publicly available.
The released models do not contain or expose raw audio, transcripts, or identifiable user information, and are provided solely to support reproducibility and further academic research.
Releasing model parameters without distributing the underlying audio data does not constitute data redistribution and is consistent with common practice in speech and language research.

\paragraph{Diversity and representativeness.}
The annotated dataset includes a diverse range of expressive speech.
More than one quarter of the annotated samples are produced by female speakers, and the data cover a wide variety of expressive and contextual settings.
While the dataset focuses on Mandarin speech, this composition reflects a deliberate effort to mitigate representational bias within the targeted linguistic and narrative domain.

\paragraph{Environmental and safety considerations.}
This work focuses on the evaluation of expressive appropriateness in speech and does not involve the deployment of generative systems in real-world settings.
As such, it does not introduce direct safety, security, or environmental risks. Nevertheless, as with other automatic speech evaluation frameworks, there is a potential risk of misuse if the proposed metrics or models are applied in high-stakes decision-making scenarios without appropriate human oversight.
We emphasize that CEAEval is intended as a research benchmark and analysis tool, rather than a standalone decision-making system.

\section*{Acknowledgments}
This work was supported by the National Natural Science Foundation of China (No. U23B2053). This work was also supported by Tencent and the Tencent-NTU Joint Research Laboratory (CENTURY), Nanyang Technological University, Singapore.


\bibliography{custom}
\newpage

\appendix

\section{Data Annotation and Inter-Annotator Reliability}
\label{appendixdataanno}

\begin{figure}[H]
  \centering
  \includegraphics[width=0.48\textwidth]{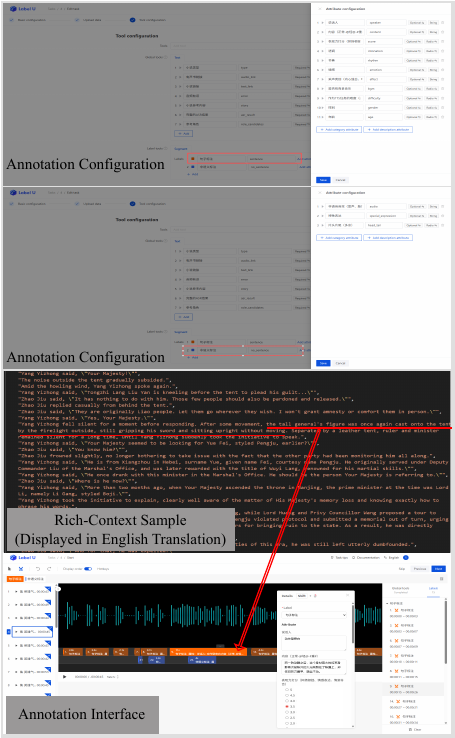}
  \caption{Annotation interface and configuration used in this work. The figure shows the annotation configuration panels, an example of the rich narrative context presented to annotators, and the annotation interface.}
  \label{fig:labelshow}
\end{figure}

We recruit 18 native Mandarin-speaking graduate students with backgrounds in speech emotion and speech perception research to annotate 16.1 hours of speech data following unified annotation guidelines.
The annotator group consists of 11 male and 7 female participants.
Before annotation, all annotators receive detailed instructions on the annotation task and interface, illustrated in Figure~\ref{fig:labelshow}, Table~\ref{tab:expressive_score}, and Table~\ref{tab:tts_difficulty}.
Annotators are informed that the data are used solely for scientific research purposes.
Each annotator labels 4 to 5 stories, and the complete annotation process takes approximately two weeks per participant.
The annotation interface supports synchronized playback of speech and its surrounding textual context, enabling annotators to consider discourse-level narrative context when making judgments.

\begin{table}[H]
\vspace{0.2cm}
\centering
\small
\setlength{\tabcolsep}{6pt}
\begin{tabular}{c m{0.8\linewidth}}
\toprule
Score & Description \\
\midrule
0--1 &
Clearly inappropriate. The expressive realization is severely misaligned with the narrative context.
Emotion, prosody, or paralinguistic behavior is clearly incorrect or contradictory, seriously disrupting comprehension or immersion. \\
\midrule
1--2 &
Weak or unnatural. Partial expressive intent is present, but major mismatches remain.
Problems such as inappropriate emotional tone, unnatural intonation or rhythm, or conflicting expressive cues limit contextual appropriateness. \\
\midrule
2--3 &
Somewhat appropriate with issues. The overall expressive direction is roughly correct, but inconsistencies in emotion, prosodic realization, or expressive emphasis reduce coherence with the narrative context. \\
\midrule
3--4 &
Generally appropriate with room for improvement. Expressive realization largely matches the narrative context and communicative intent.
Minor issues in emotional nuance, timing, or prosodic smoothness remain. \\
\midrule
4--5 &
Highly appropriate, natural, and expressive. Emotion, prosody, and paralinguistic cues are well coordinated and fully aligned with the narrative context, resulting in a fluent and convincing expressive realization. \\
\bottomrule
\end{tabular}
\caption{Scoring criteria for expressive appropriateness.}
\label{tab:expressive_score}
\end{table}

\begin{table}[htbp]
\centering
\small
\setlength{\tabcolsep}{6pt}
\begin{tabular}{c m{0.8\linewidth}}
\toprule
Score & Description \\
\midrule
0--1 &
Very easy. The utterance requires minimal expressive variation and neutral delivery. Correct rendering can be easily achieved. \\
\midrule
1--2 &
Easy. Limited expressive control is required, such as slight emphasis or mild emotional coloring. Overall delivery remains straightforward. \\
\midrule
2--3 &
Moderate difficulty. The utterance involves noticeable expressive elements, such as clear emotional cues or prosodic variation. Careful but manageable control is required. \\
\midrule
3--4 &
Difficult. The utterance demands precise expressive modulation, including nuanced emotion, timing, or intonation. Achieving a natural rendering is challenging. \\
\midrule
4--5 &
Very difficult. The utterance requires complex and fine-grained expressive control, such as subtle emotional shifts, layered prosody, or strong context dependence. It is hard to render naturally. \\
\bottomrule
\end{tabular}
\caption{Scoring criteria for TTS difficulty.}
\vspace{-0.2cm}
\label{tab:tts_difficulty}
\end{table}

\paragraph{Annotation dimensions and their relation to expressive appropriateness.}
Expressive appropriateness is annotated as an integrated perceptual judgment rather than as a deterministic combination of isolated attributes.
Annotators are instructed to assess whether the expressive realization of a speech utterance appropriately reflects the communicative intent implied by its contextual narrative.
Following established principles in Chinese broadcast speech and reading aesthetics \cite{zhang2003chinese}, several expressive attributes are annotated to provide structured support for this holistic judgment.
As summarized in Table~\ref{tab:15_dimensions}, the dataset provides fine-grained annotations across 15 distinct dimensions, covering appropriateness, prosody, emotion, text, speaker metadata, and environmental factors.

\begin{table}[!htbp]
\centering
\footnotesize 
\renewcommand{\arraystretch}{0.9} 
\begin{tabular}{@{} p{0.35\columnwidth} p{0.60\columnwidth} @{}}
\toprule
\textbf{Category} & \textbf{Annotation Dimensions} \\ \midrule

\multirow{2}{=}{\textbf{Perceptual Judgment}} 
& 1. Overall Expressive Score \\ 
& 2. TTS Difficulty \\ \midrule

\multirow{2}{=}{\textbf{Acoustic \&\newline Prosody}} 
& 3. Intonation \\
& 4. Rhythm \\ \midrule

\multirow{2}{=}{\textbf{Emotion \&\newline Intent}} 
& 5. Emotion  \\
& 6. Paralinguistic Vocalizations \\ \midrule

\multirow{3}{=}{\textbf{Context \&\newline Text}} 
& 7. Refined Textual Context \\
& 8. Refined Textual Content \\
& 9. Utterance Boundaries \\ \midrule

\multirow{3}{=}{\textbf{Speaker\newline Metadata}} 
& 10. Speaker Role Name \\
& 11. Speaker Age \\
& 12. Speaker Gender \\ \midrule

\multirow{3}{=}{\textbf{Environment}} 
& 13. Recording Conditions \\
& 14. Background Music Presence \\
& 15. Sound Events \\
\bottomrule
\end{tabular}
\caption{Overview of the 15 annotation dimensions in CEAEval-D.}
\label{tab:15_dimensions}
\end{table}

Emotional expression is annotated using open-ended textual descriptions.
Annotators are allowed to freely describe perceived emotions (e.g., happy, angry, sad) as well as compound or dynamic emotional states (e.g., calm turning into excitement), reflecting the continuous and evolving nature of expressive affect in narrative speech.
Prosodic realization is characterized along two dimensions, intonation and rhythm, following the taxonomy in \cite{zhang2003chinese}.
Intonation is categorized into four types: flat, rising, curved, and falling, capturing overall pitch movement patterns.
Rhythm is categorized into six types: brisk, heavy, low-paced, high-energy, relaxed, and tense, reflecting differences in speech tempo, energy, and stress distribution.
Recording conditions are annotated using open-ended textual descriptions to capture perceptual factors such as far-field or telephone that may influence expressive perception.
Paralinguistic vocalizations and sound events are also annotated in free-form text, covering non-verbal cues such as laughter, gasps, sighs, breath noises, or other expressive sounds.
In addition, TTS difficulty is annotated to indicate the degree of expressive control required to render an utterance appropriately under its context. Unlike expressive appropriateness, which reflects perceptual outcome, TTS difficulty captures expressive complexity from a production perspective.

\paragraph{Calibration and reliability analysis.}
Before large-scale annotation, we conduct a calibration phase in which all annotators independently label the same 14.8-minute subset of data under identical guidelines. Centralized feedback is provided to align annotators’ interpretations of the scoring criteria and reduce subjective variability. Inter-annotator agreement statistics computed on this calibrated subset are summarized in Table~\ref{tab:agreement}.

For continuous annotations, including expressive appropriateness scores, TTS difficulty, and emotion ratings, we measure inter-annotator reliability using ICC(2,1) \cite{mcgraw1996forming}, which captures absolute agreement under a two-way random-effects model.
Emotion agreement is computed by mapping categorical emotion descriptions into the Valence--Arousal--Dominance (VAD) space \cite{mohammad2025nrc}, computing ICC separately for each dimension, and averaging the results.
For categorical attributes such as intonation, rhythm, speaker age, background music presence, and speaker gender, we report percent agreement \cite{mchugh2012interrater} based on the majority label.
For paralinguistic vocalizations annotated in free-form text, agreement is quantified using an embedding-based semantic similarity \cite{reimers-2019-sentence-bert} measure, defined as the average pairwise cosine similarity among annotators’ textual descriptions.

Overall, the agreement scores indicate a high level of consistency across annotation dimensions.
Expressive appropriateness scoring achieves an \textbf{ICC of 0.87}, and emotion annotations exhibit an average \textbf{ICC of 0.93} in VAD space.
Most categorical attributes exceed \textbf{0.9 in percent agreement}, demonstrating that the annotation interface, training procedure, and calibration protocol together support reliable multi-dimensional annotation for context-rich expressive appropriateness evaluation.

\begin{table}[htbp]
\centering
\small
\setlength{\tabcolsep}{4pt}
\begin{tabular}{llcc}
\toprule
Type & Annotation & Metric & Value $\uparrow$ \\
\midrule
\multirow{3}{*}{Numeric}
& Expr. App. Score & ICC(2,1) & 0.867 \\
& TTS Difficulty & ICC(2,1) & 0.810 \\
& Emotion (VAD) & ICC(2,1) & 0.934 \\
\midrule
\multirow{5}{*}{Categorical}
& Intonation & Pct. Agr. & 0.831 \\
& Rhythm & Pct. Agr. & 0.915 \\
& Age & Pct. Agr. & 0.981 \\
& BGM & Pct. Agr. & 0.990 \\
& Gender & Pct. Agr. & 0.994 \\
\midrule
Textual
& Recording Cond. & Agreement & 0.990 \\
Textual
& Paraling. Vocal. & Agreement & 0.907 \\
\bottomrule
\end{tabular}
\caption{Inter-annotator agreement on a 14.8-minute calibration set annotated by 18 annotators. ICC(2,1) is reported for numeric annotations, percent agreement (Pct. Agr.) for closed-set categorical annotations, and embedding-based agreement for textual annotations.}
\label{tab:agreement}
\end{table}

\section{Context Construction and Context Size}
\label{app:context}

We construct a local context window composed of multiple neighboring dialogue or narrative lines.
The context is represented as an ordered list of text lines, where each item corresponds to a single utterance or narration segment in the surrounding story, as shown in Figure~\ref{fig:labelshow}.
The target line itself is treated separately from the contextual input.

The parameter context size (CTS) specifies the number of surrounding context lines provided in addition to the target line.
When CTS is set to $0$, the input consists solely of the target line, corresponding to the context-free setting.
For CTS $> 0$, the context contains exactly $C$ neighboring text lines, with a preference for lines immediately preceding the target line.
If sufficient preceding lines are available, the context includes the $C$ lines immediately before the target line.
When the target line appears near the beginning of a dialogue or narrative and fewer than $C$ preceding lines exist, the context window is expanded forward to include subsequent lines, ensuring that the total number of context lines remains fixed at $C$.
By varying the context size, we control the amount of discourse-level information available to the evaluation model, ranging from isolated utterance evaluation (CTS=0) to richer narrative-level conditioning.

\section{Contextual Prompting and Voting Strategy for the Expressive Planner}
\label{appendixplanner}

This section describes the contextual prompting and voting strategy used by the expressive planner.
\paragraph{Contextual prompting.} For each target utterance, the planner predicts an ideal expressive profile conditioned on the textual narrative context.
For a given target line, multiple planner inputs are constructed by varying the context size (CTS) from 1 to 15. Each CTS configuration, together with the same target line, is provided as an independent input to the planner using a fixed prompt template as below:

\begin{tcolorbox}[
    colback=gray!10,
    colframe=gray!50,
    title=\textbf{System Instruction: Expressive Planning from Narrative Context},
    breakable,
    left=5pt,
    right=5pt,
    top=5pt,
    bottom=5pt
]
You are an expert in speech expressiveness analysis for narrative-driven speech.
Your task is to infer the ideal expressive realization of a target utterance
based solely on the provided textual narrative context.

Given the surrounding narrative context and the target utterance,
analyze the implied communicative intent, character state, and discourse role,
and predict how the utterance should be expressed in speech.

Please consider:

1. Narrative progression, character relationships, and situational context.

2. Implied emotional state and possible emotional shifts.

3. Expressive delivery style and recording condition, including speaking distance,
   inner monologue, and sound-related delivery effects
   (e.g., phone speech, distant speech, intermittent effects).

[Input]

Narrative Context:
\%s

Target Utterance:
\%s

[Output Requirements]

Output exactly one expressive plan in the following JSON format.

The fields emotion and recording condition should be described in concise natural language (free-form), while rhythm and intonation should be selected from the predefined categories.

\{

  "emotion": "<free-form description, e.g., calm, angry, frightened, tender, conflicted>",
  
  "rhythm": "brisk / heavy / low-paced / high-energy / relaxed / tense",
  
  "intonation": "flat / rising / curved / falling",
  
  "recording condition": "<free\-form description, e.g., normal speaking, phone speech, distant voice, inner monologue>"
  
\}

Do not include any explanation or additional text outside the JSON object.
\end{tcolorbox}

\paragraph{Voting strategy.}
For each target utterance, the planner produces 15 expressive plans.
Voting is performed over the joint expressive plan rather than over individual attributes.
Specifically, each output is treated as a four-element combination of emotion, rhythm, intonation, and recording condition.
Identical combinations predicted under different context spans are grouped and counted.
The final expressive plan is selected as the combination with the highest frequency across all context variants.
In the event of a tie, the plan predicted under the longest context span is selected.
This strategy favors expressive plans that are stable across varying amounts of narrative context.

\section{Simplified Prompt with Expressive Planner}
\label{app:planprompt}

Under our framework, the system prompt for expressive appropriateness evaluation is substantially simplified by conditioning the judge model on the output of the expressive planner.
Rather than requiring the model to infer the expected expressive intent directly from a long narrative context, the planner provides a structured description of the ideal expressive realization for the target utterance.
The judge then focuses on comparing the actual speech signal against this planned expressiveness.
During both training and inference, the judge can be prompted either to directly output a score (w/o CoT) or to generate an explicit reasoning process before producing the final score (w/ CoT), enabling controlled evaluation of the effect of chain-of-thought reasoning.

\begin{tcolorbox}[
    colback=gray!10,
    colframe=gray!50,
    title=\textbf{Simplified system prompt conditioned on expressive planner output},
    breakable,
    left=5pt,
    right=5pt,
    top=5pt,
    bottom=5pt
]
You are given a speech segment and an expressive plan that describes
how this utterance should ideally be expressed in context.

Your task is to evaluate whether the actual speech realization
matches the planned expressiveness.

Please analyze the speech with respect to the following aspects:
emotion, rhythm, intonation, paralinguistic vocalizations,
recording condition, and overall naturalness.

Assign a single expressive appropriateness score between 0.0 and 5.0.

\# Scoring Guidelines:

0--1: Clearly mismatched with the expressive plan.  

1--2: Weak, unnatural, or largely monotone expression.  

2--3: Roughly matches the plan but with noticeable issues.  

3--4: Generally appropriate with minor flaws.  

4--5: Highly appropriate, vivid, and natural.

\# Ideal Expressive Plan:

emotion: \{ideal\_emotion\}  

rhythm: \{ideal\_rhythm\}  

intonation: \{ideal\_intonation\}  

recording\_condition: \{ideal\_recording\_condition\}

\% w/o CoT setting:

\% Give the final expressive appropriateness score only.

\% w/ CoT setting:

\% Think step by step and explain your analysis before giving the final score.
\end{tcolorbox}

\section{Chain-of-Thought Generation Prompt}
\label{app:cot}

Before training, we generate CoT supervision using GPT-4o, conditioned on ground-truth expressive scores, manually annotated expressive attributes, and the voted outputs of the expressive planner.
The final score is provided as a condition, and the model is instructed to generate a reasoning process that explains expressive alignment and mismatch leading to this score.
The prompt used for CoT generation is shown below.

\begin{tcolorbox}[
    colback=gray!10,
    colframe=gray!50,
    title=\textbf{Prompt for CoT generation conditioned on ground-truth score},
    breakable,
    left=5pt,
    right=5pt,
    top=5pt,
    bottom=5pt
]
You are given a spoken utterance together with its expressive evaluation result.
Your task is to generate a clear and coherent chain-of-thought that explains
why the speech receives the given expressive appropriateness score.

The final score is already provided and should be used only as a condition.
Do NOT predict, restate, or emphasize the score itself.

In your reasoning, compare the ideal expressive intent with the actual speech
realization, and analyze expressive alignment or mismatch from the following
perspectives:
emotion, rhythm, intonation, recording condition, and any relevant
paralinguistic or sound-related effects.

After comparing the two performances along each dimension, assign a score for that dimension.

\# Scoring Explanation:

0--1: Clearly mismatched with the ideal expression.  

1--2: Weak, unnatural, or largely monotone expression.  

2--3: Roughly matches the ideal expression but with noticeable issues.  

3--4: Generally appropriate with minor flaws.  

4--5: Highly appropriate, vivid, and natural.

[Input]

Target utterance:
\{target\_line\}

Ideal expressive plan:

emotion=\{ideal\_emotion\}

rhythm=\{ideal\_rhythm\}

intonation=\{ideal\_intonation\}

recording\_condition=\{ideal\_record\_condition\}

Actual expressive attributes (human annotation):

emotion=\{actual\_emotion\} 

rhythm=\{actual\_rhythm\}  

intonation=\{actual\_intonation\}  

recording\_condition=\{actual\_record\_condition\}

Ground-truth expressive score:
\{actual\_score\}

(Optional) Paralinguistic or sound-related events:
\{linguistic\_sounds\}

[Output]

Generate a natural-language chain-of-thought that explains the expressive appropriateness of the speech under the given context, explicitly providing a score for each expressive dimension.
Do not output the final score.
\end{tcolorbox}
Here, the ideal expressive attributes are provided by the expressive planner and aggregated via voting.
By conditioning CoT generation on both ideal and actual expressive attributes, the model is encouraged to explicitly reason about expressive consistency and deviation across dimensions. The generated CoT texts are translated into Chinese and used together with the original texts in bilingual training.

\section{Audio Attention Bias}
\label{appendixbias}

As discussed in the main text, the proposed audio attention bias mechanism
modulates attention strength according to the different token regions, with the goal of mitigating text-dominant reasoning under
CoT-style supervision.
This section details the construction of the region masks
\( M_{\mathrm{p}} \), \( M_{\mathrm{a}} \),
\( M_{\mathrm{CoT}} \), and \( M_{\mathrm{base}} \) in Eq.~(\ref{eqbias}),
as well as their dynamic activation during autoregressive inference.

\paragraph{Sequence structure and region annotation.}
Following the sequence formulation illustrated in Figure~\ref{fig:main},
the input to the Speech-LLM is augmented with special boundary tokens that
explicitly mark semantically distinct regions:
\[
\begin{aligned}
[ \;
& P_0, \ldots, P_n,\;
\langle a \rangle, S_0, \ldots, S_m, \langle /a \rangle,\;
\langle bos \rangle, \\
& \langle t \rangle, T_0, \ldots,
\langle f \rangle, T_i, \ldots, \langle /f \rangle, \ldots,
\langle /t \rangle, \\
& \langle s \rangle, 3.0, \langle /s \rangle
\; ]
\end{aligned}
\]
where \(P_0, \ldots, P_n\) denote system prompt tokens, \( S_0, \ldots, S_m \) denote audio tokens enclosed by \(\langle a \rangle\) and \(\langle /a \rangle\), and \( T_0, \ldots \) denote chain-of-thought tokens enclosed by \(\langle t \rangle\) and \(\langle /t \rangle\). Within the chain-of-thought region, the token pair \(\langle f \rangle\) and \(\langle /f \rangle\) marks expressive focus spans explicitly refer to speech-dependent attributes (e.g., actual emotion, intonation, or paralinguistic cues) and therefore require increased attention to the audio modality. The score value is enclosed by \(\langle s \rangle\) and \(\langle /s \rangle\) to indicate the score prediction stage.

\paragraph{Region masks.}
Based on the above sequence structure, we define four mutually exclusive binary region masks according to the explicit boundary tokens.
The system prompt mask \( M_{\mathrm{p}} \) covers the system prompt region
\( [P_0, \ldots, P_n] \).
The audio mask \( M_{\mathrm{a}} \) corresponds exclusively to the audio token region
\( [\langle a \rangle, S_0, \ldots, S_m, \langle /a \rangle] \).
The CoT mask \( M_{\mathrm{CoT}} \) marks the reasoning region
\( [\langle t \rangle, T_0, \ldots, \langle /t \rangle] \),
including all internal expressive analysis tokens.
All remaining tokens are assigned to the base mask \( M_{\mathrm{base}} \).

\paragraph{Dynamic bias activation.}
During autoregressive inference, region-specific attention bias terms in
Eq.~(\ref{eqbias}) are activated according to the boundary tokens encountered in the input sequence.
When the model enters a region marked by a start token
(e.g., \(\langle f \rangle\) or \(\langle t \rangle\)),
the corresponding bias component becomes effective for subsequent positions.
When the matching end token is reached, the bias is deactivated and attention reverts to the base setting governed by \( M_{\mathrm{base}} \).
The magnitude of each bias component is dynamically predicted from the
current hidden representation via the learnable projections
\( f_{\mathrm{p}} \), \( f_{\mathrm{a}} \), and \( f_{\mathrm{CoT}} \),
allowing the model to adapt attention strength based on contextual needs.

\begin{figure}[htbp]
  \centering
  \includegraphics[width=0.49\textwidth]{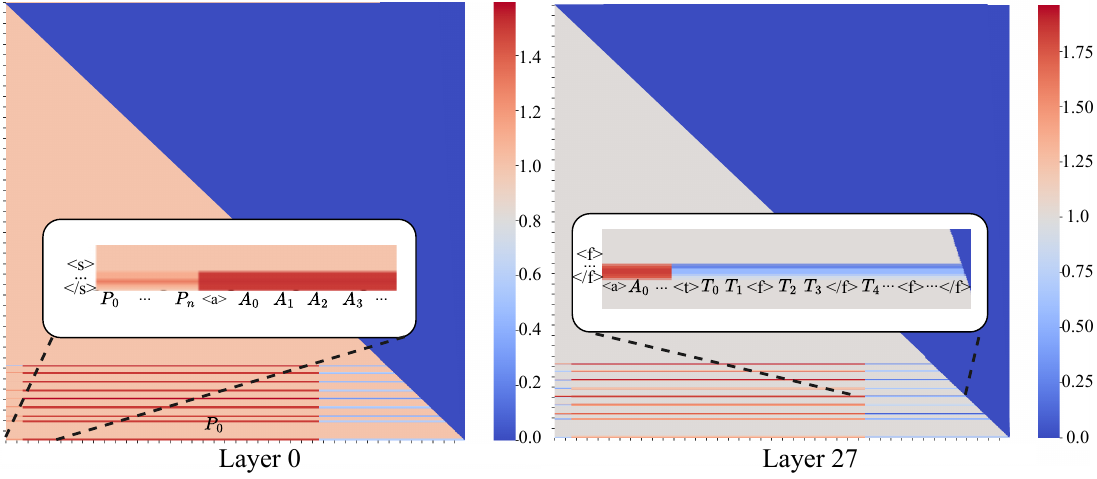}
  \vspace{-0.7cm}
  \caption{Visualization of 0th and 27th transformer layer attention bias matrices during inference.}
  \label{fig:b_show}
\end{figure}

Figure~\ref{fig:b_show} visualizes the resulting attention bias matrices at
different transformer layers under CoT-style inference, illustrating how
the proposed mechanism dynamically rebalances attention toward audio-related
regions during expressive appropriateness scoring.

\section{Filtered Training Set for Reinforcement Learning}
\label{appendix:filter}

For reinforcement learning optimization, we apply a filtering and resampling strategy to improve training stability.
Speech samples shorter than 1 second or longer than 45 seconds are removed to exclude unreliable or extreme-duration cases.
To mitigate score imbalance during policy optimization, we further perform score-balanced resampling, where samples are grouped into integer score bins and resampled to ensure approximately uniform bin frequencies.
This strategy reduces variance in reward estimation and stabilizes GRPO training.

\section{Multilingual System Prompts for Baselines}
\label{app:prompts}

To ensure reproducibility, we provide the English system prompt used for baseline expressive appropriateness evaluation below.
The Chinese prompt follows the same structure, scoring criteria, and output format, and is obtained via a direct sentence-level translation of the English version.

\begin{tcolorbox}[
    colback=gray!10,
    colframe=gray!50,
    title=\textbf{English system prompt for expressive appropriateness evaluation},
    breakable,
    left=5pt,
    right=5pt,
    top=5pt,
    bottom=5pt
]
You are given a speech segment, its surrounding textual context,
and a target utterance.

Your task is to evaluate the expressive appropriateness of the speech,
that is, whether the expressive realization of the utterance appropriately matches the intended communicative setting implied by the context.

Please consider the following aspects in an integrated manner:
emotion, rhythm, intonation, recording condition, and any relevant paralinguistic or sound-related cues.

Assign a single expressive appropriateness score in the range of 0.0 to 5.0.

\# Scoring Guidelines:

0--1: Clearly inappropriate; expressive realization is severely mismatched with the context.

1--2: Weak or unnatural; major expressive issues remain.

2--3: Somewhat appropriate with noticeable problems.

3--4: Generally appropriate with minor flaws.

4--5: Highly appropriate, natural, and expressive.

\# Output Format:

Return only the final score in the following format:
<score>1.5</score>

\# Reasoning Mode (optional):

The model may either directly output the score (w/o CoT),
or generate an explicit reasoning process before the final score (w/ CoT),
depending on the experimental setting.

[Context]  

\{context\}

[Target Utterance]  

\{target line\}
\end{tcolorbox}

\section{Model Parameter Counts}
\label{app:params}
To facilitate a clearer comparison of model capacity as discussed in Section~\ref{sec:experiments}, we summarize the parameter counts of all evaluated models in Table~\ref{tab:model_params}. Note that for the planner-assisted baselines, the 8B parameters of the expressive planner (Qwen3-8B) are effectively utilized during inference.

\begin{table}[h]
\centering
\begin{tabular}{lc}
\toprule
\textbf{Model} & \textbf{Model Size} \\
\midrule
Qwen2.5-Omni & 7B \\
Kimi-Audio & 7B \\
Phi-4-MM & 5.6B \\
Gemma-3n & 6B \\
Step-Audio-R1 & 32B \\
Midashenglm & 7B \\
GPT-4o-Audio & - \\
Gemini-1.5-Pro & - \\
Voxtral-Mini & 3B \\
Qwen3-Omni & 30B \\
\midrule
Ours & 7B (Judge) + 8B (Planner) \\
\bottomrule
\end{tabular}
\caption{Parameter counts of the models evaluated in this work.}
\label{tab:model_params}
\end{table}

\section{Case Study: Planner and Judge Output}
\label{app:case_study}

To illustrate how CEAEval-M performs context-rich expressive appropriateness evaluation in practice, we present a representative case study below. The example demonstrates how the text-only Expressive Planner infers the ideal expressive profile from a multi-turn narrative context, and how the Speech-LLM Judge leverages this profile to perform step-by-step chain-of-thought (CoT) reasoning before predicting the final score.

\begin{tcolorbox}[
    colback=gray!5,
    colframe=gray!50,
    title=\textbf{Case Study: Context-Rich Evaluation},
    breakable,
    left=5pt,
    right=5pt,
    top=5pt,
    bottom=5pt
]
\textbf{[Input Narrative Context (CTS = 4)]}
\begin{itemize}[leftmargin=*, itemsep=2pt, parsep=0pt]
    \item \textit{Line 1:} Lin Nianlei said: ``Ah, it's really warm.''
    \item \textit{Line 2:} Qin Yu said nothing more and continued walking forward at a steady pace. Wearing gloves, Lin Nianlei brushed her hands over the icicles hanging from the windowsills of the houses, then suddenly shouted.
    \item \textit{Line 3:} Lin Nianlei said: ``You big dummy.''
    \item \textit{Line 4:} Qin Yu said: ``What?''
\end{itemize}

\textbf{[Target Utterance (Line 5)]} \\
\textit{Lin Nianlei said: ``Cheer up. What's already happened can't be changed by anyone. But since we live in such a time, we should always believe that tomorrow will be beautiful.''}

\vspace{0.2cm}
\textbf{[Input Speech]} \\
\begin{center}
\vspace{-0.4cm}
  \fbox{\parbox{0.95\textwidth}{\centering  \small \texttt{target\_utterance\_speech.wav}  }}
  \vspace{0.2cm}
\end{center}

\textbf{[Expressive Planner Output]}
\begin{lstlisting}
The ideal performance in this context is: { "emotion": "gentle", "rhythm": "soothing", "intonation": "rising intonation", "recording condition": "normal speech" }
\end{lstlisting}

\textbf{[Speech-LLM Judge Output (CoT + Score)]} \\
The ideal emotion is gentleness, while the actual performance conveys comfort, encouragement, and calmness. The emotional direction is consistent and naturally expressive, resulting in an emotion score of 4.2. In terms of rhythm, the ideal style is soothing, while the actual delivery is light and lively, with smooth pacing and well-placed pauses, yielding a rhythm score of 4.0. The ideal intonation is rising, while the actual intonation is curved. Although the intonation pattern does not exactly match, it is rich and appropriate to the context, resulting in an intonation score of 4.0. The ideal recording condition is normal speech, and the actual performance is also normal speech, with good overall clarity, leading to a recording condition score of 4.0. In addition, inhalation sounds appear in the speech and are naturally integrated into the context, further enhancing emotional expression. Overall, the performance is natural and fluent, with minor shortcomings in detail. I assign a final score of <score>4.0</score>. 

\hrulefill

\vspace{0.2cm}
\textbf{[Human Expressive Appropriateness Score]} \\
Ground Truth: 4.0
\end{tcolorbox}

\end{document}